\input harvmac

\def\p{\partial}

\def\half{{1\over 2}}

\def\barl{\bar{\lambda}}

\Title{hep-th/0011170}{\vbox{\centerline{Quantum Corrections to Noncommutative
Solitons}}}
\centerline{ Miao Li }
\centerline{\it Institute for Theoretical Physics}
\centerline{\it Academia Sinica, Beijing}
\centerline{And} 

\centerline{\it Department of Physics}
\centerline{\it National Taiwan University}
\centerline{\it Taipei 106, Taiwan} 
\centerline{\tt mli@phys.ntu.edu.tw}

\vskip1.5cm

Noncommutative solitons are easier to find in a noncommutative field theory.
Similarly, the one-loop 
quantum corrections to the mass of a noncommutative soliton are easier to
compute, in a real scalar
theory in $2+1$ dimensions. We carry out this computation in this paper.
We also discuss the model with a double-well 
potential, and conjecture that there is a partial symmetry restoration in a
vacuum state.

\vskip0.5cm
\Date{November 2000}

\nref\gms{R. Gopakumar, S. Minwalla and A. Strominger,
``Noncommutative Solitons,''
JHEP {\bf 0005} (2000) 020, hep-th/0003160.}
\nref\omm{E. Brezin, C. Itzykson, G. Parisi and J. Zuber, Comm.
Math. Phys. 59 (1978) 35.}
\nref\mm{I.R. Klebanov, ``String theory in two dimensions,"
hep-th/9108019; P. Di Francesco, P. Ginsparg and J. Zinn-Justin,
``2D gravity and random matrices," hep-th/9306153.} 
\nref\gs{S.S. Gubser and S.L. Sondhi, ``Phase structure 
of non-commutative scalar field theories," hep-th/0006119.}
\nref\qft{see for instance, J. Zinn-Justin, ``Quantum Field Theory and 
Critical Phenomena,"  Oxford, UK: Clarendon (1989).}
\nref\mrs{S. Minwalla, M. Van Raamsdonk, N. Seiberg, 
``Noncommutative Perturbative Dynamics," hep-th/9912072.} 
\nref\tachy{A. Sen,``SO(32) spinors of type I and other solitons
on brane-antibrane pair," JHEP 9809 (1998) 023, hep-th/9808141;
``Stable non-BPS bound states of BPS D-branes," JHEP 9808 (1998) 010,
hep-th/9805019; ``Tachyon condensation on the brane antibrane system,"
JHEP 9808 (1998) 012, hep-th/9805170; ``BPS D-branes on non-supersymmetric
cycles," JHEP 9812 (1998) 021, hep-th/9812031; ``Descent relations
among bosonic D-branes," Int. J. Mod. Phys. A14 (1999) 4061, hep-th/9902105.}
\nref\kth{E. Witten, ``D-branes and K-theory," JHEP 9812 (1998) 019,
hep-th/9810188; P. Horava, ``Type IIA D-branes, K-theory, and matrix theory,"
Adv. Theor. Math. Phys. 2 (1999) 1373, hep-th/9812135.}
\nref\piljin{P. Yi, ``Membranes from Five-Branes and Fundamental Strings from Dp 
Branes," Nucl.Phys. B550 (1999) 214,  hep-th/9901159;
O. Bergman, K. Hori and P. Yi, ``Confinement on the Brane,"  Nucl.Phys. B580 (2000) 289,
hep-th/0002223.}
\nref\wit{E. Witten, ``Overview Of K-Theory Applied To Strings,"
hep-th/0007175.}
\nref\dmr{K. Dasgupta, S. Mukhi and G. Rajesh,
``Noncommutative Tachyons,''
JHEP {\bf 0006} (2000) 022, hep-th/0005006.}
\nref\hklm{
J. A. Harvey, P. Kraus, F. Larsen and E. J. Martinec,
``D-branes and Strings as Non-commutative Solitons,''
JHEP {\bf 0007} (2000) 042, hep-th/0005031.}
\nref\ew{E. Witten,
``Noncommutative Tachyons And String Field Theory,''
hep-th/0006071.}
\nref\seib{N. Seiberg, ``A Note on Background Independence 
in Noncommutative Gauge Theories, Matrix Model
and Tachyon Condensation," hep-th/0008013.}
\nref\gmas{R. Gopakumar, S. Minwalla and  Strominger,
``Symmetry Restoration and Tachyon Condensation in Open String 
Theory,"  hep-th/0007226.}
\nref\krs{P. Kraus, A. Rajaraman and S. Shenker, ``Tachyon
Condensation in Noncommutative Gauge Theory," hep-th/0010016.} 
\nref\miao{M. Li, ``Note on noncommutative tachyon in matrix models,"
hep-th/0010058.}
\nref\ha{H. Awata, S. Hirano and Y. Hyakutake, ``Tachyon Condensation 
and Graviton Production in Matrix Theory," hep-th/9902158.}
\nref\hkl{J.A. Harvey, P. Kraus and F. Larsen, ``Exact 
Noncommutative Solitons,"  
hep-th/0010060.}
\nref\more{D. Bak, ``Exact Solutions of Multi-Vortices and False 
Vacuum Bubbles in Noncommutative Abelian-Higgs Theories,"  
hep-th/0008204;
D. Gross and N. Nekrasov,  ``Solitons in 
Noncommutative Gauge Theory," hep-th/0010090;
D. Bak, K. Lee, J.-H. Park, ``Noncommutative Vortex Solitons,"
hep-th/0011099.}
\nref\sol{D. Gross and N. Nekrasov, ``Monopoles and Strings 
in Noncommutative Gauge Theory,"  hep-th/0005204; 
``Dynamics of Strings in Noncommutative Gauge Theory,"
hep-th/0007204; 
A.P. Polychronakos, ``Flux tube solutions 
in noncommutative gauge theories," hep-th/0007043.
D. P. Jatkar, G. Mandal and S. R. Wadia, 
``Nielsen-Olesen Vortices in Noncommutative Abelian Higgs Model,"
hep-th/0007078.}
\nref\bss{T.Banks, N. Seiberg and S. Shenker,
``Branes from Matrices," Nucl.Phys. B490 (1997) 91, 
hep-th/9612157.} 
\nref\miaol{M. Li, ``Strings from IIB matrices," Nucl. Phys.
B499 (1997) 149, hep-th/9612222.}
\nref\aw{L. Alvarez-Gaume and S. Wadia, ``Gauge Theory on a Quantum Phase Space,"
hep-th/0006219. } 
\nref\radu{R. Tatar, ``A Note on Non-Commutative Field Theory 
and Stability of Brane-Antibrane Systems,"  hep-th/0009213;
G. Mandal, S. R. Wadia, ``Matrix Model, Noncommutative Gauge 
Theory and the Tachyon Potential," hep-th/0011144.}

\newsec{Introduction}

By now there is a fairly large literature on the subject of noncommutative 
solitons \gms\ in field theories as well as in string theory. However, there is 
little study on the quantum properties of these solitons. We will take
first steps in this direction.

Noncommutativity in a sense drastically simplifies the task of the search 
for solitons. Noncommutativity is the cause of UV/IR connection, and drastically
reduces the number of degrees of freedom. Technically, the field theory
is essentially replaced by an one dimensional matrix model in the large
noncommutativity limit.  And the {\it essential} dynamic degrees of freedom
are the eigen-values of the matrix. A soliton may be called a single 
eigen-value soliton.  In the same vein, noncommutativity also simplifies the
computation of quantum corrections to the spectrum of solitons.
The reason is that in the large $\theta$ limit, one can integrate
out the angular variables in the unitary matrix, in the decomposition
$\phi =U\Lambda U^+$, the result is a modification of the wave functions
by the Vandermonde determinant with the Hamiltonian remaining unchanged.
To be more accurate, excitations of angular variables will cause 
quantum corrections suppressed by $g^2$, the coupling constant in question.
Thus the problem of computing quantum corrections to the spectrum of 
solitons boils down to a simple quantum mechanical problem. This is to
be contrasted to the problem of computing one-loop correction to
a conventional soliton in a usual quantum field theory, where ingenious
techniques are often required to carry out the calculations.
In a noncommutative field theory, solving the differential equation
is reduced to solving an algebraic equation, and computing a functional
determinant is reduced to computing a number. We will
carry out the ``one-loop'' computation in the next section. We will
treat the spatial kinetic term as a perturbation, since it is suppressed
by $1/\theta$.  

The same reason prompts one to suspect that the folklore of symmetry 
breaking is no longer valid in a noncommutative field theory. We will
conjecture in sect.3 that indeed in a $2+1$ dimensional noncommutative 
scalar field theory, symmetry is partially restored in a ground state in 
the large $\theta$ limit. We will provide some evidence to this effect. 
For finitely many eigenvalues, the $Z_2$ symmetry with a double-well
potential is not broken. In the spacetime picture, this corresponds to
the partial symmetry restoration in a finite region whose size
depends on $\theta$. This phase breaks translational invariance thus
deserves further study.

\newsec{One-loop corrections}

We will mostly consider a noncommutative real scalar in $2+1$ dimensions.
The action is
\eqn\sact{S={1\over 2\pi g^2\theta}\int dtd^2x\left(\half(\p_t\phi)^2-\half
(\p_i\phi)^2-V(\phi)\right),}
where all products are understood as the star product, $V(\phi)$ is a
polynomial in $\phi$, and is assumed to be bounded from below.
The coupling constant $g^2\theta$ is chosen to contain a factor
$\theta$, the noncommutative parameter appearing in
\eqn\commu{[x_1,x_2]=i\theta.}
This choice is purely for later convenience.
We define the coupling constant in such a way that the leading power
in $V(\phi)$ has a dimensionless numerical coefficient. For instance,
in the case of a $\phi^4$ theory, $V(\phi)=1/4 \phi^4+\dots$.
The dimension of $g^2\theta$ can be determined as follows. First, from
the kinetic terms, one finds
$$[g^2\theta]=[\phi^2]L.$$
If the leading power of $V(\phi)$ is $n$, then by comparing this
leading term with the kinetic term, one finds 
$$[\phi]=L^{-{2\over n-2}}.$$
Of course in order to have an interacting theory and solitons,
$n>2$. Since the dimension of $\theta$ is $L^2$, finally we have
\eqn\sdim{[g^2]=L^{-{n+2\over n-2}}.}
For $n=4$, $[g^2]=M^3$, and for $n=6$, $[g^2]=M^2$. Without
noncommutativity, the former theory is super renormalizable,
and th latter is marginally renormalizable.

To discuss the GMS solitons, it is convenient to rescale
the spatial coordinates $x_i\rightarrow \sqrt{\theta}x_i$
Furthermore, it is useful to replace the integration over
$x_i$ by the trace using $\int d^2x=2\pi \tr$ after the
rescaling. The Hamiltonian is written in a neat form
\eqn\hamil{H={1\over g^2}\tr \left(\half (\p_t\phi)^2
-{1\over 2\theta}[x_i,\phi]^2 +V(\phi)\right).}
In the large $\theta$ limit, we see that we can drop the
spatial derivative terms $[x_i,\phi]^2$.
The above Hamiltonian describes a matrix model, one that
is a slight modification of the old familiar $c=1$
matrix model. To cast the Hamiltonian in the more familiar
matrix form, introduce the creation and the annihilation operators
\eqn\caop{a={1\over \sqrt{2}}(x_1+ix_2),\quad a^+={1\over\sqrt{2}}
(x_1-ix_2)}
satisfying $[a,a^+]=1$. Field $\phi$ is operator valued,
thus can be written in the form
\eqn\matrixs{\phi=\sum_{m,n}\phi_{mn}|m\rangle\langle n|,}
where $|m\rangle$ are the normalized eigen-states of the
number operator $a^+a$. Apparently, the multiplication
operation becomes that of matrices. 
The Hamiltonian \hamil\ does not have a $U(\infty)$ gauge
symmetry. However, if one throws away the spatial kinetic
term, which is small in the large $\theta$ limit, the
model possesses a global $U(\infty)$ symmetry. 

In the large $\theta$ limit, one can always diagonalize
$\phi$:
\eqn\diago{\phi=\sum_m \lambda_m |m\rangle\langle m|,}
and the Hamiltonian becomes simply a sum of infinitely
many decoupled terms
\eqn\dech{H_0={1\over g^2}\sum_m \left(\half (\p_t\lambda_m)^2
+V(\lambda_m)\right).}
For a static configuration, to minimize the energy, we need to minimize 
all the individual terms, so that
\eqn\minie{{dV(\lambda_m)\over d\lambda_m}=0.}
Thus the static dynamics is dictated by the local
minima of the potential $V$. We shall always assume the
global minimum of $V$ be $0$ in this section, so the vacuum state has a 
zero energy classically. GMS observed that, if there exists
another local minimum $\barl$ with $V(\barl)$, then there
exist solitons $\phi=\barl|m\rangle\langle m|$ and multi-solutions
as suppositions of these solitons. Although it appears
that these results are rather trivial from the matrix model
perspective, the solitons are nontrivial as they are truly
lumps in the two dimensional space. Translated into function
of $x_i$, $|m\rangle\langle m|$ is a polynomial with a Gaussian
damping factor.

It is the special feature of a noncommutative field theory
that the supposition of all single eigen-value solitons
give the false vacuum, since
\eqn\supps{\phi =\sum \barl |m\rangle\langle m| =\barl,}
namely, although each soliton is well-localized around
$x_i=0$, their supposition gives rise to a constant configuration
of the scalar field.

For a single soliton, the classical energy is degenerate 
regardless which eigen-value $\lambda_m$ is excited to
$\barl$:
\eqn\ecal{E_m={1\over g^2}V(\barl).}
This degeneracy is lifted by turning on the spatial
kinetic energy. For the diagonal configuration \diago,
this term $-(1/g^2\theta)\tr [a,\phi][a^+,\phi]$ reads
\eqn\skin{{1\over g^2\theta}\sum_m m (\lambda_m-
\lambda_{m-1})^2.}
Consider the more general situation when all $\lambda_m$
are dynamic, namely when they depend on time, the
Hamiltonian is
\eqn\thamil{H={1\over g^2}\sum_m\left(\half
(\p_t\lambda_m)^2+{1\over\theta}m(\lambda_m-
\lambda_{m-1})^2+V(\lambda_m)\right).}
Assume the global minimum of $V$ is at $\lambda=0$,
then to the first order in $1/\theta$, the m-th
soliton has an energy
\eqn\mener{E_m={1\over g^2}\left(V(\barl)+{2m+1\over\theta}\barl^2
\right).}
The degeneracy is lifted by a small term suppressed
by $1/\theta$. We computed the above correction by
simply substituting the unmodified soliton into
\thamil. The exact static solution when the
spatial kinetic energy is turned on is different
from the simple ansatz $\lambda_m=\barl$,
$\lambda_n=0$, $n\neq m$. However, this modification
does not change the result \mener. To see this,
we can expand the Hamiltonian \thamil\  around
the unmodified soliton configuration and find that
the correction to $\phi$ is suppressed by $1/\theta$,
all terms in the expansion except for the one
showing up in \mener\ are then suppressed by $1/\theta^2$
upon substitution of the modified solution.

To discuss the quantum corrections to the spectrum of solitons,
again we will ignore the spatial kinetic term first.
Without the presence of this term, it is well-known
in the old matrix models that one can integrate
out angular variables first. Decompose the full matrix
as written in \matrixs\ into $\phi=U\Lambda U^+$,
where $\Lambda$ is the full diagonal form as in \diago,
and $U$ is a unitary operator $U\in U(\infty)$. To see
how integrating out $U$ affects the resulting theory,
it is better to work with the path integral:
\eqn\pathi{\int [d\Lambda][dU]\exp(iS).}
Now if $S$ contains only the time kinetic term and the
potential term, $U$ can be explicitly integrated
out \refs{\omm,\mm}. Consider the propagator between
states with fixed initial $\Lambda_i$ and fixed final
$\Lambda_f$, the path integral yields
\eqn\pathin{\int [d\Lambda] \Delta (\Lambda_i)
\Delta^{-1} (\Lambda_f) \exp(iS(\Lambda)),}
where $S(\Lambda)$ is the action by substituting
$\phi=\Lambda$ into $S(\phi)$, and $\Delta(\Lambda)$
is the Vandermonde determinant
\eqn\vand{\Delta (\Lambda)=\prod_{m<n}(\lambda_m-\lambda_n).}
Note that in \pathin\ we integrated out $U_i$
but not $U_f$, otherwise we would get a factor $\Delta
(\Lambda_f)$ instead of $\Delta^{-1}(\Lambda_f)$.
Result \pathin\ can be obtained by put the matrix model
on a discrete time lattice, integration of angular
variables yields many Vandermonde determinants, and
all the intermediate determinants cancel.
We thus see that integrating out the
angular variable $U$ merely modifies the wave function
$\Psi (\Lambda)\rightarrow \Delta (\Lambda)\Psi (\Lambda)$
without changing the Hamiltonian \dech.  
This antisymmetric
factors comes from the phase space of the angular variables.

When the initial wave function is a nontrivial function of
these variables, integrating
out them is actually more subtle than
described above. In any case, if one uses the arguments in
the first reference in \mm, one will see that excitations
of angular variables bring in corrections only at the
order $g^2$, one-loop higher than what we will be interested
in this paper. All these angular variables are flat directions.
In particular, the mode corresponding to the center of
the soliton is a collection of these variables (the $U$
matrix is simply $\exp(i(\alpha a+\bar{\alpha}a^+))$). Since
the mass of the soliton is proportional to $g^{-2}$, the
kinetic term of the soliton is roughly $(g^2/\lambda^2)p_\alpha^2$,
we see that indeed this term also brings in correction at
the order $g^2$, as it should be the case in general \foot{
The problem of treating the collective coordinates is always
subtle in a commutative field theory, we see that in the 
noncommutative case, it is relatively simple, although 
the center degree of freedom $\alpha$ is composed of infinitely
many angles and requires a careful treatment.}.
By freezing
angular directions except taking the phase space into
account, we also demand the soliton stay at rest.

It becomes important to remember that there is a spatial
kinetic term in the full action breaking the $U(\infty)$
symmetry. Without this term, we would start with a totally
symmetric wave function $\Psi(\Lambda)$ and end up with 
a totally asymmetric wave function $\Delta (\Lambda)\Psi
(\Lambda)$, and each eigen-value $\lambda_m$ is transmutated
into a fermion, as in the old matrix models. In this case
the vacuum cannot be $\lambda_m=0$, and there are no soliton
solutions at all! Thus, even though the classical perturbation
introduced by the spatial kinetic term is suppressed by
$1/\theta$ as in \thamil, its quantum correction is enormous
to change the vacuum structure. 

We will first ignore the spatial kinetic term in discussing the
quantum corrections to the spectrum of solitons, and consider
its effects later. Since there is no restriction on the wave
function, all eigen-values are completely decoupled, as the
Hamiltonian \dech\ clearly indicates. The canonical momentum of
$\lambda_m$ computed from \dech\ is
\eqn\cmom{p_m={1\over g^2}\p_t \lambda_m.}
With this identification, the Hamiltonian becomes
\eqn\chamil{H=\sum_m \left({g^2\over 2}p_m^2+{1\over g^2}
V(\lambda_m)\right).}
Clearly, if $\lambda_m=0$ is the absolute minimum point of
$V(\lambda_m)$, then the ground state $\Psi_0(\lambda_m)$
is localized around this minimum, and the ground state of
the whole system is given by $\Psi_0(\Lambda)=\prod_m
\psi_0(\lambda_m)$. The classical soliton with a $\lambda_m
=\barl$ is unstable against quantum tunneling. We will not
be interested in this tunneling in this article. What we are
interested in is the perturbative quantum corrections to
the classical energy of the soliton, thus the full perturbative
energy is a Taylor series in $g^2$. While the quantum tunneling
is suppressed by a factor $\exp(-c/g^2)$ with the constant
$c$ depending on the details of the potential $V$.

The first order correction to the energy is of order $O(g^0)$
and is normally called the one-loop correction. While it is often
a quite technical problem to compute this one-loop correction
in a usual quantum field theory, the computation is almost trivial
in the noncommutative field theory without $U(\infty)$ symmetry.
We need only expand $V(\lambda_m)$ around $\barl$ to the
second order in $\Delta \lambda_m=\lambda_m-\barl$. 
The Hamiltonian for $\lambda_m$ to this order reads
\eqn\mhamil{H_m={1\over g^2}V(\barl)+{g^2\over 2}p_m^2
+{1\over 2g^2}V''(\barl)(\Delta\lambda_m)^2,}
where $V''(\barl)$ is the second derivative of $V$ at $\barl$
and is positive.
Thus the wave function for $\lambda_m$ is that of the ground
state of a harmonic oscillator, it is just
\eqn\gwave{\Psi(\Delta\lambda_m)=\exp(-{\sqrt{V''(\barl)}\over 2g^2}
(\Delta\lambda_m)^2).}
More generally, the n-th excited state has an energy
\eqn\exener{E={1\over g^2}V(\barl)+(n+\half)\sqrt{V''(\barl)}.}
We might think this is the quantum corrected energy of the
soliton with $n$ quanta bound to it. This is not true, we need
to take the energy of vacuum into account. When all eigen-values
stay in their ground state, the classical energy vanishes.
Quantum mechanically, to the first order in $g^2$, there is a 
correction
\eqn\cener{\half \sqrt{V''(0)}}
to the energy of a single eigen-value. The sum of these corrections 
diverges as usual
in a field theory. For a soliton excited to its n-th level,
all other eigen-values still stay in their ground state around
$\lambda=0$ except $\lambda_m$, thus the subtracted energy of the
soliton is
\eqn\mthe{E_m={1\over g^2}V(\barl)+\half (\sqrt{V''(\barl)}
-\sqrt{V''(0)})+n\sqrt{V''(\barl)}.}
The above is the true quantum corrected energy of a soliton bound
to n quanta. When $n=0$, we have the one-loop corrected energy
of the soliton. The condition for the one-loop correction
to be much smaller than the classical result is
\eqn\compa{g^2\ll {V(\barl)\over |\sqrt{V''(\barl)}
-\sqrt{V''(0)}|}.}

Next we consider the correction brought about by the spatial 
kinetic term. When this term is present, it is impossible to
integrate out the angular variables $U$ as before. However,
since this term is suppressed by $1/\theta$, it is reasonable to
assume that integrating out $U$ will result a correction to
the wave function by $\Delta (\Lambda)$ and a small correction
to the Hamiltonian. We will ignore this correction to the 
Hamiltonian.  We have seen in \mener\ that the classical correction
induced by the spatial kinetic term is suppressed by a factor
$1/\theta$, and not surprisingly, we will see that the quantum
correction induced by this term is further suppressed by
a factor $g^2$. Denote $H=H_0+\Delta H$, where $H_0$ is the
Hamiltonian as in \dech, and $\Delta H$ is the spatial kinetic
term. Let $\Psi_i$ be the eigenstates of $H_0$, using the
standard perturbation theory, the eigen-values of $H$ is determined
by the following equation
\eqn\pert{\det[\langle \Psi_i|\Delta H|\Psi_j\rangle
-\delta_{ij}(E-E_i)]=0.}
Consider $\Psi_i$ to be the wave functions when all $\lambda_n$
$n\ne m$ are excited around $\lambda=0$, and the m-th
eigen-value $\lambda_m$ is excited around $\barl$. Since 
the energy-levels $E_i$ are not degenerate, it is rather
easy to execute the perturbation calculation. We first assume
that the off-diagonal elements $\langle \Psi_i|\Delta H|
\Psi_j\rangle$ can be ignored, then the correction to 
$E_m$ is simply
\eqn\corr{\langle \Psi_m |\Delta H|\Psi_m\rangle
= {1\over g^2\theta}\sum _n (2n+1)\langle\Psi_m |
\lambda^2_n|\Psi_m\rangle,}
where $\Psi_m$ is given by the product of \gwave\
and the ground state wave functions of $\lambda_n$
centering around $\lambda_n=0$. In getting the above
expressing we have observed that the expectation
value of the nearest-neighbor coupling $\lambda_n
\lambda_{n+1}$ vanishes. \corr\ is easy to compute,
we have
\eqn\comc{{1\over\theta}(m+\half){1\over\sqrt{V''
(\barl)}}+{1\over\theta}\sum'_n (n+\half)
{1\over\sqrt{V''(0)}},}
where the primed sum does not include $n=m$. The above
sum is divergent. To get a finite result, again
we need to subtract the correction to the vacuum energy,
thus
\eqn\scorr{\langle \Psi_m |\Delta H|\Psi_m\rangle
={1\over\theta}(m+\half)({1\over\sqrt{V''
(\barl)}}-{1\over \sqrt{V''(0)}}).}
Compared with the classical correction in \mener\
we see that indeed this term is further suppressed
by a factor $g^2$.

Next we shall argue that the off-diagonal elements
of $\Delta H$ can be ignored. To see this, we
use another method to get the same result as in
\scorr. The spatial kinetic term can be divided 
into two parts. One part is a sum of the terms
$${1\over g^2\theta}(2n+1)\lambda_n^2.$$
The second part is a sum of the nearest neighbor
coupling
$$-{1\over g^2\theta}n\lambda_n\lambda_{n-1}.$$
Now treat the second part as a perturbation,
while the first part is included in $H_0$, we see
that the total one-loop quantum correction
is
\eqn\qcorr{\half\sqrt{V''(\barl)+2(2m+1)/\theta}.}
Expanding the above result we get the second term
to be the one in \scorr. The higher order terms are
suppressed by more factors of $1/\theta$. Further,
it is easy to see that in the nearest neighbor
coupling, the elements of $\lambda_n$ between
the ground state and the excited states are
all vanishing except the first excited state. 
According to the perturbation theory, the correction
induced by this is proportional to
$$(\langle\Psi |1/(g^2\theta) \lambda_n\lambda_{n-1}
|\Phi\rangle)^2/(E_1-E_0),$$
where $E_1-E_0$ is the energy difference between state
$\Phi$ and state $\Psi$. Due to the appearance of
square of the off-diagonal element, this term is
suppressed by $1/\theta^2$. The off-diagonal element
is independent of $g$.

To summarize, the corrected energy of the
m-th single soliton is
\eqn\tcorr{\eqalign{E_m&={1\over g^2}(V(\barl )
+{2m+1\over\theta}\barl^2)+\half (\sqrt{V''(\barl)}
-\sqrt{V''(0)})\cr
&+{1\over\theta}(m+\half)({1\over\sqrt{V''(\barl)}}
-{1\over\sqrt{V''(0)}}).}}
If we demand the quantum correction induced by the
spatial kinetic term be much smaller than its
classical counterpart, we must impose
\eqn\scon{g^2\ll {\barl^2\sqrt{V''(\barl)V''(0)}\over
|\sqrt{V''(\barl)}-\sqrt{V''(0)}|}.}
Further, if we demand the quantum correction
induced by the spatial kinetic term be much
smaller than the one in the decoupled system, there
must be
\eqn\tcon{{1\over\theta}\ll {\sqrt{V''(\barl)V''(0)}\over
m+\half}.}
This condition may be violated for a sufficiently large
$m$. 

To have a feeling about the result \tcorr, consider a
$\phi^4$ theory. $V'(\lambda)$ is a polynomial of degree
$3$, so it has three zeros. To have a soliton, there
must be two zeros at $\lambda=0$ and $\lambda=\barl$.
The third zero must be real too, denote it by $\mu$. Thus
\eqn\alg{V'(\lambda)=\lambda (\lambda-\barl)(\lambda-
\mu).}
The relevant data for quantum corrections are
\eqn\dataq{V''(0)=\mu\barl , \quad
V''(\barl) =\barl (\barl-\mu).}
For both of them to be positive, $\mu$ and $\barl$ must
have the same sign, choose them to be positive; and
$\barl >\mu$. $\lambda=\mu$ is a local maximum of $V$.
The potential is a deformed double-well. Integrating
\alg\ to get $V$ and set $V(0)=0$, we find
\eqn\solm{V(\barl)= {1\over 12}\barl^3 (2\mu-\barl).}
This value must be positive, therefore $2\mu>\barl$.
Applying this condition to \dataq, we have
$V''(0)>V''(\barl)$. We therefore see that the
leading quantum correction in \tcorr\ is negative,
the quantum correction due to the kinetic term is
positive.

Quantum corrections to multiple solitons sitting at the
same point can be readily carried out. In this case
a few $\lambda$'s are excited to $\barl$ and there is no
interaction at the classical level, since even the
spatial kinetic term vanishes. Quantum mechanically, there
is no correction at the one loop level to the interaction
energy, since the cross terms such as $\Delta\lambda_m
\Delta\lambda_{m-1}$ has vanishing expectation value.

It is also interesting to study quantum correction to
the energy of two separated solitons. The simplest
case is a soliton localized at $x=0$, given by $\barl
|0\rangle\langle 0|$, and a soliton localized at a another
point given by $\barl |z\rangle\langle z|$. The second
soliton can not be written as a diagonal matrix, so new
technique is required.

\newsec{Partial Symmetry Restoration}

In this section we will consider the noncommutative $\phi^4$
theory in $2+1$ dimensions. We assume the potential take the 
form \eqn\php{V(\phi)={1\over 4}\phi^4-{\mu^2\over 2}\phi^2.}
According to the dimensional analysis performed in the
previous section, $\phi$ has an energy dimension, so does
$\mu$, and $[g^2]=M^3$. Classically, as in the
ordinary commutative field theory, the vacua are degenerate
at $\phi=\pm\mu$. There is no soliton unlike what was
studied in the previous section. Quantum mechanically, 
the vacua remain degenerate for the commutative field theory. 
What happens in the noncommutative case? 

We conjecture that in a true ground state,
the vacuum expectation values of the first few $\lambda_m$ are zero, 
with the largest $m$ determined roughly by $\mu^2\theta$.
The remaining infinitely many eigen-values stay in the
valley of the double-well potential. Such a vacuum seems
to break translational invariance, thus it is
something like a ``stripe phase".
We make this conjecture for the zero temperature theory,
unlike what was considered recently by Gubser and Sondhi
\gs, where an Euclidean noncommutative field theory was 
studied. 

It is well-known that for a quantum mechanical system with
the double-well potential \php, the degeneracy of two
perturbative ground states centering around $\phi=
\pm \mu$ is lifted nonperturbatively, due to the tunneling
effects. Let the Hamiltonian of this quantum mechanics be
\eqn\qhamil{H={1\over g^2}(\half (\p_t\phi)^2+V(\phi)),}
with $V(\phi)$ given in \php.
Let $E_0(g^2)$ be the ground state energy computed
by the perturbative expansion method. The nonperturbative
effects lift the degeneracy and the two eigen-values of
the Hamiltonian become $E_\pm(g^2)$, the difference 
between these two energies can be computed by the WKB method
and turns out to be of the form \qft
\eqn\ldeg{E_+(g^2)-E_-(g^2)=c_1{\mu^{5/2}\over g}e^{-{c_2\mu^3\over g^2}},}
where $c_{1,2}$ are positive dimensionless constants . 
This result is good in the limit $g^2/\mu^3\ll 1$.
Note that the true ground state energy 
$E_-$ is lower than the perturbative ground state energy
$E_0$.

In a quantum field theory with potential \php, there are
two extremal limits. In one extreme, the spatial derivative
terms can be ignored, and in this case the field theory is
ultra-local. At every spatial point, $\phi$ fluctuates
independently. If we divide space into cells with a small
volume $v$, then the effective coupling is $g^2/v$, very 
large in the UV limit.   In this limit the $\phi^4$ term
dominates the dynamics, and the degeneracy is surely
lifted in this limit, although the result \ldeg\ can no
longer be trusted. In the other extremal limit, the
spatial kinetic term is important, so $\phi$ is forced
to fluctuate collectively. Denote this zero mode by $\phi_0$,
its effective coupling is $g^2/V$ where $V$ to the infrared
cut-off on the whole volume. Since in the large volume
limit, the effective coupling is small, and the energy
gap \ldeg\ becomes accurate in this limit:
\eqn\egap{E_+-E_-=c_1{\sqrt{V}\over g}e^{-{c_2V\over g^2}}.}
In the thermodynamic limit $V=\infty$, the energy gap
tends to $0$ rapidly, the degeneracy is resumed,
the $Z_2$ symmetry is spontaneously broken.

In the realistic case, one has to study the system carefully,
taking into account the spatial kinetic energy. It turns
out for a real scalar field,
symmetry is always broken in 2 dimensions and above. For
a complex scalar, the kinetic term
does not suppress the infrared correlations of the Goldstone
boson in 2 dimensions, and there
is no symmetry breaking.

We now turn to our $2+1$ dimensional noncommutative $\phi^4$
theory. As emphasized in the previous section, it is
a difficult problem to integrate out the off-diagonal
modes of $\phi$ when the spatial kinetic term is present.
Nevertheless we assume that the correction induced by this
term is smaller than its classical value when evaluated for $\phi
=\Lambda$, a diagonal matrix. Thus the net result of integrating
out $U$ in the decomposition $\phi=U\Lambda U^+$ is
the Vandermonde determinant modifying the wave function.
The Hamiltonian is still given by \thamil. For convenience,
we write down this Hamiltonian again
\eqn\fhamil{H={1\over g^2}\sum_m\left(\half
(\p_t\lambda_m)^2+{1\over\theta}m(\lambda_m-
\lambda_{m-1})^2+{1\over 4}\lambda_m^4-\half \mu^2
\lambda_m^2\right).}
In the limit $\theta=\infty$, the spatial kinetic term drops
out, the system becomes of a collection of $\phi^4$
quantum mechanical systems, and the vacuum state is
the state when all eigen-values $\lambda_m$ stay in their
ground state of energy $E_-$, so there is no spontaneous
symmetry breaking. The first excited states
are degenerate: When all $\lambda_m$ are in their ground
state except only one eigen-value in the state with energy
$E_+$. The energy of these first excited state, after subtracted
the vacuum energy, is given by $E_+-E_-$ as in \ldeg,
and curiously, is very small when $g^2/\mu^3\ll 1$.
We conclude that in the large $\theta$ limit, there is no
symmetry breaking, and there is an energy gap which is 
nonperturbative in nature.
The fact that a $2+1$ dimensional noncommutative scalar field
theory does not exhibit spontaneous symmetry breaking
is due to the drastic reduction of effective number of 
degrees of freedom. In a way, it is similar to the ultra-local
commutative field theory. It is also different from the latter,
since the effective coupling constant is always $g^2$ and
all the eigen-values $\lambda_m$ have the same amount of
fluctuations, while in the ultra-local field theory, the 
fluctuations become violent in the UV limit, since the effective
coupling increases in the UV limit.

When the kinetic term is turned on, the tremendous
degeneracy of the first excited states is lifted. 
Hamiltonian \fhamil\ describes a chain on a half infinite
line with nearest neighbor coupling. When $\theta$ large
and $m$ small, the spatial kinetic term is not strong,
one would imagine these eigen-values essentially behave
like decoupled eigen-values, and the $Z_2$ symmetry is not
broken for them. For $m$ large enough, the spatial kinetic
term starts to play an role, we will argue that it will
derive almost all eigen-values to their minimal point.

The one dimensional chain \fhamil, although looks simple,
can not be solved exactly. We will use perturbation arguments.
Classically, since the kinetic term is positive definite,
the minimal energy is achieved when $\lambda_m=\pm \mu$.
Quantum mechanically, all $\lambda_m$ fluctuate, it is
a tricky question as to what we shall take as our unperturbed
Hamiltonian. We shall consider two situations separately.
In one hypothetical case, $Z_2$ symmetry is not broken
for all $\lambda$'s. In the other case, $Z_2$ symmetry is
broken.

Consider the assumption that the symmetry is not broken for
all eigenvalues. In this case
we will not take the Hamiltonian in \dech\
as the unperturbed one, rather, we take
\eqn\deco{H_0={1\over g^2}\sum_m \left(\half (\p_t\lambda_m)^2
+(-\half \mu^2 +{2m+1\over \theta})\lambda_m^2+{1\over 4}
\lambda_m^4\right)}
as the unperturbed Hamiltonian. The perturbation
is then the nearest neighbor coupling
\eqn\pert{H_I=-{2\over g^2\theta}\sum_m m\lambda_m
\lambda_{m-1}.}
Without perturbation \pert, the system described by \deco\ is simple.
The effective mass square for $\lambda_m$ is 
$$\mu_m^2={2(2m+1)\over\theta}-\mu^2.$$
When this parameter is positive, the potential has a minimum at
$\lambda_m=0$. In the following argument, we ignore the finitely
many eigen-values for which $\mu_m^2$ is negative, since contribution
of these eigenvalues to energy is always finite.
Now the zero-point energy for $\lambda_m$ with positive
$\mu_m$ is simply $(1/2)\mu_m$. Since all the wave 
functions of $\lambda$'s are even functions, the perturbation \pert\
has vanishing expectation value in this ``supposed to be" vacuum,
the vacuum energy is given by, ignoring higher loop corrections
\eqn\venery{E_0=\sum_{m=M}{\mu_m\over 2},}
where $M$ is the smallest $m$ making $\mu_m^2$ positive.
If we introduce a cut-off $m=N$ on the sum, then the second sum 
scales roughly as
\eqn\esti{{2\over 3\sqrt{\theta}}N^{3/2}.}

As we remarked before, if we do not take the nearest neighbor
coupling as a perturbation, then there are two classical minima 
when all $\lambda_m=\mu$ or all $\lambda=-\mu$. 
We now study the quantum fluctuation around one of them, say
$\lambda=\mu$. Each eigenvalue contributes to the
classical energy an amount$-\mu^4/(4g^2)$, and each contribute
a zero point energy $\mu/\sqrt{2}$. The sum of these contribution
diverges as
\eqn\negae{-({\mu^4\over 4g^2}-{\mu\over\sqrt{2}})N.}
We are assuming the perturbation theory is correct, so $g^2/\mu^3\ll
1$, the above contribution is negative.
Expanding the whole Hamiltonian around $\lambda_m=\mu$,
and denote $\Delta\lambda_m=\lambda_m-\mu$, the Hamiltonian,
for $m\ge M$, is given by a Gaussian part
\eqn\decou{H_0={1\over g^2}\sum_m \left(\half (\p_t\Delta\lambda_m)^2
+(\mu^2 +{2m+1\over \theta})\Delta\lambda_m^2\right),}
a perturbation
\eqn\pertu{H_I=-{2\over g^2\theta}\sum_m m\Delta\lambda_m\Delta
\lambda_{m-1},}
and a higher order sum.
The zero-point fluctuation determined by \decou\ is
\eqn\zerop{\sum_m \half \left({2(2m+1)\over \theta}+2\mu^2
\right)^{1/2}.}
This term also diverges as the sum \venery\ in the same fashion
as in \esti. But the difference between this sum and \venery\
is roughly
\eqn\dff{{3\over 4}\mu^2\sqrt{\theta}\sqrt{N}.}
Although it is positive,
in the limit $N\rightarrow\infty$, this term is overwhelmed by
the negative energy \negae, so the symmetry broken phase
$\lambda=\mu$  has much smaller energy.

For $\lambda_m$ with $\mu_m^2$ negative, the story can be
completely different. For these eigen-values, even in
the symmetric phase, the energy is negative, and if we adopt
\pert\ as perturbation, it has zero expectation value
in the symmetric state. For this term to be a small perturbation,
we require its element between the symmetric state with
energy $E_-$ and the first excited state with energy $E_+$
to be smaller than $(\Delta E\mu_m)^{1/2}$. We find the condition
\eqn\condi{{m^2\over\theta^2}\ll c_1g^3|\mu_m|^{-1/2}\exp(-{c_2
|\mu_m|^3\over g^2}),}
it can be satisfied for small $m$ and large $\theta$.

Thus, it is likely that the one-dimension chain
described by \fhamil\ has a zero-temperature phase
in which finitely many eigen-values are in the symmetric
phase. These eigen-values still have an effective
negative $\mu_m^2$, thus $m$ is roughly smaller than
$\mu^2\theta$. As pointed out in \gms, the corresponding
state $|m\rangle\langle m|$ associated with this
eigen-value has a spatial size $\sqrt{m}$, or in
terms of the original, un-rescaled coordinates, its
size is $\sqrt{m}\sqrt{\theta}$. With the condition
$m<\mu^2\theta$, we find the size be smaller than
$\mu\theta$. A more careful analysis should incorporate
a condition such as \condi. For instance, one may
require that the spatial kinetic energy between the
two neighboring eigen-values to be smaller than the 
lowering of energy due to quantum tunneling, given
in \ldeg. The former is roughly $m\mu^2/(g^2\theta)$,
for the difference $\lambda_m-\lambda_{m-1}$ is roughly
$\mu$ in a symmetric phase.
Thus, 
\eqn\condit{m< g\sqrt{\mu}e^{-{c_2\mu^3\over g^2}}\theta,}
so the physical size of symmetry restoration
\eqn\rest{\sqrt{m\theta}\sim g^{1/2}\mu^{1/4}e^{-{c_2\mu^3
\over 2g^2}}\theta=f(g,\mu)\theta.}

This result may just reflect the fact
that in noncommutative space, the fluctuation of the
field within size $f(g,\mu)\theta$ is unavoidably large
due to space-space uncertainty. This size grows
when $\theta$ is increased. There ought to be a close
relation between the partial symmetry restoration and
the UV/IR connection at the perturbative level found in
\mrs.

It goes without saying that the above result is to
be taken only as a substantiated conjecture, since we
have ignored dynamic fluctuations of angular variables, and
their effect introduced through the spatial kinetic
term. 

The partial symmetry breaking is interesting. However
it raises a puzzle: In such a phase, translational
invariance seems to be broken. In a sense  this
vacuum is similar to the stripe phase discussed in
\gs.  The simple reason that the spontaneously
translational symmetry-breaking is possible is that
the translation generators are part of the angular
variables, which are all flat directions and can
spontaneously stay at a point.
Further detailed study is desirable to clarify
how spontaneous translational symmetry-breaking occurs.
Nevertheless its occurrence is not unique. As discussed
in \refs{\krs,\miao}, tachyon condensation in a D-brane
anti-D-brane system also spontaneously breaks translational
invariance in matrix models, the condensation occurs
in a finite region at the classical level. For
other discussions of the noncommutative tachyon, see
\radu.
The cause of this phenomenon
is the opposite sign of the B field on the two
D-branes. In the system discussed here, the ``tachyon"
field does not condense in a finite region.

\newsec{Conclusion and Discussions}

We have in this paper mainly considered a noncommutative
field theory with a real scalar, in $2+1$ dimensions.
This dimensionality is interesting in string
theory, for instance in a brane-anti-brane system, 
lower dimensional branes are constructed as solitons
of co-dimension 2 \refs{\tachy-\ha}. However, to
get results applicable to string theory, the gauge field
must be introduced. Results obtained here will be
modified with the presence of the gauge field. The
most important modification will be that the complete
symmetry breaking will become possible again, for
otherwise Sen's tachyon condensation scenario would be wrong, this
is unlikely to happen. It should also be interesting to
study quantum corrections to the exact solitons
discussed recently in \refs{\hkl-\more} and other
solitons involving the gauge field \sol. 

When there is a gauge field in the system, unlike
the commutative case, a scalar can remain real and
is still coupled to the gauge field in the adjoint
representation. A tachyon on a unstable D-brane
in a type II string theory realizes this situation.
The problem of computing quantum corrections to a unstable soliton
in a gauge theory is complicated by two things in
a noncommutative gauged system. First, even though one can
still diagonalize a real scalar in the operator representation,
one is left with the problem of treating the time
component of the gauge field. Second, the
spatial derivative terms and the
Yang-Mills term are important. To see this,
recall that a covariant derivative $D_\mu\phi$ is
replaced by a commutator $[X_\mu,\phi]$ in the
matrix representation \bss\ \miaol. Rescaling of coordinates 
$x_i\rightarrow \sqrt{\theta}x_i$ is absorbed into
a rescaling of the gauge field, so the Yang-Mills
term $\tr [X_\mu,X_\mu]^2$ also receives a rescaling,
the effective Yang-Mills coupling is $g^2\theta$ rather than
$g^2$. Thus the Yang-Mills part is strongly coupled
compared to the self coupling of the scalar. In a $\phi^4$
theory, the effective dimensionless coupling is
$g^2\theta/\mu$.

To discuss gauge symmetry breaking, one must first understand
better the Higgs mechanism in this setting \aw. We expect
no surprises here.

Acknowledgments. 
I would like to thank P. M. Ho, Y. C. Kao, H. S. Yang and especially
P. Yi for discussions.
This work was supported by a grant of NSC and by a 
``Hundred People Project'' grant of Academia Sinica.

\vfill
\eject

\listrefs

\end